\documentclass{article}\newcommand{\address}[1]{\date{\medskip#1}}
\usepackage{amsthm}
\theoremstyle{plain} 
\newtheorem{theorem}{Theorem}
\newtheorem{lemma}[theorem]{Lemma}
\usepackage{prooftree}
\usepackage{pslatex}
\usepackage{fancybox}
\usepackage{amssymb,dsfont,wasysym}
\usepackage[all]{xy}\SelectTips{cm}{} \CompileMatrices
\usepackage{qsymbols}
\usepackage{pdflscape}
\usepackage{graphics}
\usepackage{alltt}
\usepackage{url}

\newcommand{\Coq}{\textsc{Coq}} 
\newcommand{\Nat}{\mathbb{N}}

\newcommand{\TcaG}{\ensuremath{\mathcal{T^{C[`a]}_G}}}

\newcommand{\mcp}{{\it muddy children puzzle}} %
\newcommand{\ini}{{\it initial event}} %
\newcommand{\pro}{{\it progression event}} %
\newcommand{\epev}{{\it epistemic event}}
\newcommand{\nc}{\textsf{c}}
\newcommand{\nmc}{\textsf{m}}
\newcommand{\mud}{`m}
\newcommand{\atleast}{`l}
\newcommand{\exact}{\varepsilon}

\begin{document}
\title{\textsf{\textbf{\large LIP Research Report RR2007-50}}\\[5pt]
Dynamic Logic of Common Knowledge \\
in a Proof Assistant}
\author{Pierre Lescanne\thanks{Corresponding author.
\texttt{email: Pierre.Lescanne@ens-lyon.fr}},\quad
J\'{e}r\^ome Puiss\'{e}gur}
\address{Universit\'{e} de Lyon, Ecole Normale Sup\'{e}rieure de Lyon, CNRS (LIP), \\
46, all\'{e}e d'Italie, 69364 Lyon 07, FRANCE}
\maketitle

\begin{abstract}
  Common Knowledge Logic is meant to describe situations of the real
  world where a group of agents is involved.  These agents share
  knowledge and make strong statements on the knowledge of the other
  agents (the so called \emph{common knowledge}).  But as we know, the
  real world changes and overall information on what is known about the
  world changes as well.  The changes are described by dynamic logic.
  To describe knowledge changes, dynamic logic should be combined with
  logic of common knowledge.  In this paper we describe experiments
  which we have made about the integration in a unique framework of
  common knowledge logic and dynamic logic in the proof assistant
  \Coq.  This results in a set of fully checked proofs for readable
  statements.  We describe the framework and how a proof can be
  conducted.
\end{abstract}

\begin{quotation}
\noindent\textbf{keywords:} Common Knowledge, Dynamic Logic, Proof Assistant
\end{quotation}

\section{Introduction}

{\it Common knowledge logic} is about the knowledge of the world,
whereas {\it dynamic logic} is about the changes of the world.  Both
are presented as {\it modal logic}.  In this paper we propose to
analyze reasoning in a combination of those logics through a
mechanization by a proof assistant.

By experience, we know that the knowledge we have of the world is not
perennial, but is meant to evolve.  Therefore, any faithful and
complete approach of reasoning of agents about their surrounding world
requires to take that evolution into account and to combine a logic
that describes the state of the knowledge at a given time and a logic
that accounts the changes due to external events.  This kind of work
is known as \emph{belief revision} (or knowledge revision in our case)
and is advocated by Johan van
Benthem~\cite{benthem01:_games_dynam_epist_logic}.  In this paper,
following the work
of~\cite{baltag99,baltag98,dynamics96:_johan_benth,benthem01:_games_dynam_epist_logic,ditmarsch05:_dynam},
we combine two logics: the first logic is common knowledge
logic~\cite{geanakoplos94:_handb_game_theor,806735,806736,mccarthy77:_model_theor_knowl}
and the second one is dynamic
logic~\cite{harel79:_first_order_dynam_logic,harel00:_dynam_logic}.
The combination of both is called \emph{dynamic logic of common
  knowledge}.  The idea is not new but the novelty is that we do that combination in a
proof assistant.

As we are neither designers of modal logic, nor philosophers, but only
proof assistant users, what is presented in this paper is not a
general discussion on the interest or the advantage of combining
logics or how this can be made more appropriately.  What we present is
a record of experiments done on a mechanization of dynamic logic of
common knowledge in \Coq, one of the proof assistants available on the
market.  By the use of higher logic and mechanization this activity
sheds light on the reality of reasoning in dynamic logic of common
knowledge and on how the two components, namely epistemic and dynamic
fit together. This paper does not address any comparison on using one
proof assistant or another in that kind of implementation exercise.
We feel that actually higher order proof assistants like
ACL-2~\cite{555902}, HOL~\cite{team05:_hol_system_descr},
Isabelle~\cite{Nipkow-Paulson-Wenzel:2002}, LEGO~\cite{legoWWW},
PHoX~\cite{raffalli05:_phox_proof_assis} or PVS~\cite{PVS}, are not so
deeply different w.r.t. modal logic and that such a comparison would
not be informative for the reader.  We prefer to focus on the
experience itself, hoping that what has been learned will help
designers of logics.  We have taken \Coq, because we practiced
it~\cite{lescanne06:_mechan_coq_AMAI} and we have an expert environment
around us.  This paper is essentially a careful examination of what is
necessary to make an actual proof of correctness.  We have chosen the
\mcp{} (again not a very original choice) and we introduce the reader
to the \Coq{} script.

\subsubsection*{Why experiences on a proof assistant?}

We noticed that most of the presentations about logic of common
knowledge or dynamic logic or a combination of both were made either
through a model approach where no specific care is given to actual
deductions, with rules and axioms\footnote{A notable exception related
  to our approach is the formulation of linear temporal logic in COQ
  done by Solange
  Coupet-Grimal~\cite{coupet-grimal03:_axiom_of_linear_logic}.  Her
  development is a shallow embedding when ours is a deep one.}.  When
proofs are given they are done at an intermediary level of
abstraction, whereas we advocate a deep level, where no detail is left
over.  We are typically at a \emph{proof theory} level.  With a proof
theoretic background, we feel that proofs and deductions are of main
importance as it has been shown with most of experience with proof
assistants.  To summarize, this paper is about the actual integration
of common knowledge and dynamic logic in a unique framework in a proof
assistant.  It relies on a previous work by the first
author~\cite{lescanne06:_mechan_coq_AMAI} and is associated with two
scripts:

\centerline{\footnotesize\url{http://perso.ens-lyon.fr/pierre.lescanne/COQ/EpistemicLogic.v8}}

and

\centerline{\footnotesize\url{http://perso.ens-lyon.fr/pierre.lescanne/COQ/EpistemicAndDynamicLogic.v}.}


\section{Dynamic logic of common knowledge}

\subsubsection*{Common knowledge logic} 

Common knowledge logic is a modal logic with two main modalities.  One
modality~$K_i$, which is associated with each agent $i$, is the
\emph{knowledge modality}.  It is meant to express the knowledge an
agent has on statements, facts and propositions.  For instance,
$K_i(`v)$ reads as \emph{$i$~knows $`v$}.  The modality $C_G$, which
is associated with a group $G$ of agents is the \emph{common knowledge
  modality}.  $C_G(`v)$ translates the fact that a knowledge is common
to a group~$G$ of agents, not only each agent in the group $G$ knows
$`v$, but also he knows that the others know $`v$ and he knows that
the others know that the others know $`v$, and this recursively.
$C_G(`v)$ reads as $`v$ \emph{is a common knowledge of the group} $G$.
It is formalized as a fixed point by an axiom and a rule:

\[    \prooftree
  \justifies "|-" C_G \varphi "->" \varphi \wedge E_G C_G \varphi \using FixPoint_C
  \endprooftree
  \qquad \qquad 
  \prooftree
  "|-" \rho "->" \varphi \wedge E_G \rho
  \justifies "|-" \rho "->" C_G \varphi \using GreatestFixPoint_C
  \endprooftree
\]

\subsubsection*{Dynamic logic} 

Dynamic logic makes events modalities.  There are as many modalities
as there are events.  If $`a$ is an event, then $[`a]$ is a modality
and one writes $[`a]`v$ the proposition modified by an event $`a$.
If an universe satisfies $`v$, after the event $`a$ has been
performed on it, the transformed universe satisfies $[`a]`v$.

\subsubsection*{Hilbert-style}

Hilbert-style is what has been chosen in the COQ implementation.  
 It is convenient both from the point of view of
its presentation and from the point of view of its mechanization in a
proof assistant.  Therefore the forthcoming rules and axioms will be
presented in that framework.

The reason why one cannot use a natural deduction of a sequent
calculus approach is essentially due to the Generalization Rule.  If
one accepts such a rule in natural deduction, one gets
\begin{center}
  \(\prooftree `G"|-" `v \justifies K_i(`G) "|-"
  K_i(`v)\endprooftree\)
\end{center}
This requires to extend the operator $K_i$ to contexts like $`G$.  If
instead of $K_i$ one uses a modality $\Box$, one says that $\Box(`G)$
is a \emph{``boxed context''}.  Actually \emph{linear logic}~\cite{LL}
is perhaps the archetypal modal logic and the equivalent of $K_i$ is
the modality \emph{of course} written ``!''.  The equivalent of
Generalization Rule is a rule called also \emph{of course}.  Without
that rule the proof net presentation is somewhat
simple~\cite{lafont95from}.  Its introduction requires a machinery of
boxes which increases its complexity.  See~\cite{AvronHMP98} for a discussion.

\subsubsection*{The axioms}

The axioms of modal logic are those of {\it classical logic} plus two
axioms and one rule for each modality $M$:

\begin{itemize}
\item \emph{Normalization axiom K$_M$:} $"|-" M \varphi "->" M (\varphi "->" `j) "->" M `j$
\item  \emph{Necessitation axiom T$_M$:} $"|-" M \varphi "->" \varphi$
\item \emph{Generalization rule Gen$_M$:} $ \prooftree "|-" \varphi
  \justifies "|-" M \varphi \endprooftree $
\end{itemize}

These axioms of modal logic have to be duplicated for dynamic logic and
common knowledge logic.

\subsection{Epistemic and dynamic modalities: purely epistemic
  propositions}

The central issue of this paper is to show how to integrate common
knowledge and dynamic logics in a unique framework for using in a
proof assistant.  First we define a logic that we call \TcaG{} (see
Figure~\ref{fig:logic}). An interesting feature of \TcaG{} is axiom
{\bf KT1}:
\[\forall `v: proposition ~ \forall`a: event ~ \forall i`:G,
\quad "|-" K_i[`a]`v "->" [`a]K_i`v\] %
It is well known in epistemic-temporal logic~\cite{FaginHMV95} and is
appropriate for dynamic logic of common knowledge.  It reads
\emph{``if agent $i$ knows that, after event $`a$, $\varphi$ holds,
  then one can infer that, after event $`a$, agent $i$ knows that
  $\varphi$ holds''}.  This axiom allows commuting epistemic and
dynamic modalities in one direction.  Note that the converse is quite
dubious in natural language and would certainly be rejected by
philosophers.  Indeed if \emph{after $`a$, I~know that $`v$ holds},
because event $`a$ is precisely to let me know proposition $`v$, then
there no reason to infer that \emph{I know that $`v$ has to hold after
  $`a$}.  But looking carefully at axiom {\bf KT1}, one notices that
event $`a$ is transforming not actually the world in its physical
reality, but the knowledge the agent has of it.  Therefore to avoid
troubles and paradoxes, we consider only events $`a$ that are {\it
  ``purely epistemic''}.  This means that in our approach of dynamic
logic of common knowledge, we consider only actions or events that
change the perception of the world which agents have, not the world
itself.  We borrowed this concept of purely epistemic event from A.
Baltag~\cite{baltag98,baltag99}.

\subsection{The axiomatization of dynamic logic of common knowledge}

  \begin{figure}[t]
\centering
\doublebox{
\begin{math}
  \begin{array}[c]{c}
    \begin{array}[c]{c@{\qquad}c}
    \prooftree "|-"_{\mathcal K} \varphi \justifies "|-" \varphi
    \using Classical
    \endprooftree
    &
    \prooftree "|-" \varphi~~~~ "|-" \varphi "->" `j
    \justifies "|-" `j \using MP
    \endprooftree
  \end{array}
  \\[12pt]
  \begin{array}[c]{c@{\qquad}c@{\qquad}c}
    \prooftree
  \justifies "|-" K_i \varphi "->" K_i (\varphi "->" `j) "->" K_i `j \using K_K
  \endprooftree
  &
  \prooftree
  \justifies "|-" K_i \varphi "->" \varphi \using T_K
  \endprooftree
  &
  \prooftree
  "|-" \varphi
  \justifies "|-" K_i \varphi \using Gen_K
  \endprooftree
  \end{array}
  \\[12pt]
  \prooftree
  \justifies "|-" E_G\varphi \leftrightarrow \bigwedge_{i\in G} K_i \varphi \using Def_E
  \endprooftree
  \\[12pt]
  \begin{array}[c]{c@{\qquad}c}
    \prooftree
  \justifies "|-" C_G \varphi "->" \varphi \wedge E_G C_G \varphi \using FixPoint_C
  \endprooftree
  &
  \prooftree
  "|-" \rho "->" \varphi \wedge E_G \rho
  \justifies "|-" \rho "->" C_G \varphi \using GreatestFixPoint_C
  \endprooftree
\end{array}
  \\[12pt]
  \begin{array}[c]{c@{\qquad}cc}
      \prooftree
  \justifies "|-" [`a] \varphi "->" [`a] (\varphi "->" `j) "->" [`a] `j \using K_{[`a]}
  \endprooftree
  &
  \prooftree
  \justifies "|-" [`a] \varphi "->" \varphi \using T_{[`a]}
  \endprooftree
  &
  \prooftree
  "|-" \varphi
  \justifies "|-" [`a] \varphi\using Gen_{[`a]}
  \endprooftree
  \end{array}
  \\[12pt]
  \prooftree
  \justifies "|-" K_i [`a] \varphi "->" [`a] K_i \varphi \using KT1
  \endprooftree
\end{array}
\end{math}     
\centering
}

\caption{The dynamic logic of common knowledge $\mathcal T^{C[`a]}_G$}
\label{fig:logic}
\end{figure}

For the common knowledge modality $C_G$ we have chosen the
axiomatization proposed and implemented in Coq by the first of
us~\cite{lescanne06:_mechan_coq_AMAI}.  The whole dynamic logic of common
knowledge is made of the following ingredients:
\begin{itemize}
\item the logic $\mathbb{T}$ for $K$ and for $[`a]$,
\item the definition of \emph{shared knowledge} $E_G$,
\item the definition of \emph{common knowledge} $C_G$ by a 
  fixpoint axiom and a rule that says that it is the greatest fixpoint, 
\item the axiom \textbf{KT1} that makes the connection between dynamic
  logic and common knowledge logic.
\end{itemize}


\section{A running example: the \mcp}

The \mcp{} will serve as an example to show how dynamic and knowledge
logic have been integrated in \Coq.  This problem is presented by
several
authors~\cite{FaginHMV95,geanakoplos94:_handb_game_theor,meyer95:epist_logic}
as an illustration of common knowledge logic. The problem considers
amazing children who are be able to carry perfectly logical reasoning.

\subsection{The statement}

First, let us recall the puzzle.  The reader who knows the puzzle can
skip this part and jump to Section~\ref{sec:GainCom}, collecting the
axioms.  We follow more or less the presentation of Meyer and van der
Hoek~\cite{meyer95:epist_logic}.

$n+1$ children are standing in a circle around their father. There are
$\nmc+1$ ($\nmc \in \{0,...,\nc\}$) children with mud on their face.  The
children can see each other, but they cannot see themselves. In
particular, they do not know if they have mud on their face.  Father
says aloud: ``There is at least one child with mud on its face.'' Then
he asks: ``Will all children who know they have mud on their face
please step forward?'' This procedure is repeated until, after the
$\nmc+1$-th time Father has asked the same question, all muddy children
miraculously step forward.

The conclusion which happens eventually is the result of a logical
reasoning made by the children, especially the muddy ones, about what
they know initially and what they know about the changes on what they
know.  It is a perfect example of an common knowledge and dynamic
reasoning which fits with our frameworks.

\subsection{The formalization}

In this section, we try to say what justified our statements. A reader
interested only by the formal rules and the mechanized reasoning can
jump over the text and go directly to the formal statements.  This
discussion is interesting to understand why we have chosen this system
of axioms.

\subsubsection*{Two events}

In this puzzle, the action are not very elaborated, since after
Father's first statement, he keeps repeating the same sentence.
Therefore we consider two events, one that starts the scenario and
that we write ``\currency'', it is also called the \ini{}, and one that
corresponds to the sentence Father repeats and that we will write
``*'', it is also called the \pro.  In our dynamic logic of common
knowledge, we will have two types of propositions: $[\textrm{\currency}]`v$
and~$[*]`v$.  We will also write $[*]^k`v$ for $[*]...[*]`v$ where
$[*]$ is repeated $\nmc$ times.  Clearly $[*]^0`v$ means $`v$.  In
\Coq{}, we will use the identifiers \texttt{Point} (abbreviated
in \texttt{[\currency]} in \Coq) and \texttt{Star} (abbreviated
in \texttt{[*]} in \Coq).

\subsubsection*{Definitions}

To study this puzzle, we must describe formally the situation and so define basic properties with axioms.

Let $\nc\in\mathbb N$ and $\nmc \in \{0,...,\nc\}$, so that $\nc+1$ is the number
of children (there is at least one of them) and $\nmc+1$ the number of
muddy ones (there is also at least one of them).  Let~$G$ be the group
of all children, of cardinality $\nc+1$: we identify it with
$\{1,...,\nc+1\}$.

Let $\mud_i$ ($i \in \{1,...,\nc+1\}$) be the proposition ``child $i$ has
mud on his face''.

Let $\atleast_j$ ($j \in \mathbb N$) be the proposition ``at least $j$ children have mud on their face''.

Let $\exact_j$ ($j \in \mathbb N$) be the proposition ``exactly $j$
children have mud on their face'', which is defined as follows:
\[
\mathbf{EQ}_{\atleast\exact}: \qquad \forall j \in \Nat,\quad "|-" \exact_j \leftrightarrow \atleast_j \wedge \neg \atleast_{j+1}
\]
what one can read ``there are exactly $j$ muddy children if and only
if there are at least $j$ and at the most $j$ ones''. Two trivial
properties can be proved from this axiom (the proof is made in the
\Coq~file): first, ``if there are at least but not exactly $j$ muddy
children, then there are at least $j+1$ ones'', which is:
\[\mathbf{IMP}_{\atleast\exact} \qquad \forall j \in \Nat,\quad "|-" \atleast_j
\wedge \neg \exact_j "->" \atleast_{j+1} 
\]
secondly, a principle of exclusion, ``there cannot be exactly $j$ and
at least $j+1$ muddy children'', which is: 
\[
\mathbf{EXCLU}_{\atleast\exact} \qquad \forall j \in \mathbb N, \quad "|-"
\neg ( \atleast_{j+1} \wedge \exact_j) 
\]

These propositions describe the \emph{``physical world''}, i.e., the
physical state of the children, whether they are muddy or not.  They
form the type \emph{physical proposition}.  As we only take into
account {\it epistemic events}, physical propositions are {\it
  ``persistent''}, which means they are not modified by {\it epistemic
  events}.  This property is axiomatized as follows:
\[ \mathbf{PERSIST} \qquad \forall p: physical~proposition ~~ \forall
`a: epistemic~event, \quad "|-" p "->" [`a] p 
\]

\subsubsection*{The \ini~and its consequences}

First, \emph{Father says loudly that there is at least one muddy
  child}: therefore this proposition becomes common knowledge. If
\verb+TRUE+ is the logical constant, we notice that it is the only
``true'' proposition available to the children initially. The effect
of the first statement is as follows:
\[
\mathbf{MC1}_1 \qquad "|-" [\texttt{\currency}]\verb+TRUE+ "->" C_G \atleast_1
\]
this is the first axiom of our formalization.

\emph{The children are not blind}, they see each other and they get
pieces of information from it. The \ini{} records what they get:
\emph{every child counts the number of muddy children in front of
  him/her}.  In particular, the muddy ones see $\nmc$ muddy children,
thus they get a knowledge about the total number of muddy children,
namely $\nmc$ or $\nmc+1$:
\[
\mathbf{MC1}_2 \qquad \forall i \in G,\quad "|-" [\texttt{\currency}] \verb+TRUE+ "->" \mud_i "->" K_i(\exact_{\nmc} \vee \exact_{\nmc+1})
\]

Defined that way, the \ini{} is an \epev: No further action will
change the world, only the knowledge the agents own on the world will evolve.
Therefore the muddy children problem is a paradigmatic example.

We said that physical propositions are persistent, but they are not
the only ones.  Indeed, the muddy children are able to remember what
they have seen initially, in other words, the part $\mud_i "->" K_i(\exact_{\nmc}
\vee \exact_{\nmc+1})$ of axiom {\bf MC1$_2$} is also persistent: 
\begin{center}
  \begin{math}
    \textbf{PERS}_{MC1_2} ~~ \forall `a: event ~ \forall i \in G, "|-"
    (\mud_i "->" K_i(\exact_{\nmc} \vee \exact_{\nmc+1})) "->" [`a]
    (\mud_i "->" K_i(\exact_{\nmc} \vee \exact_{\nmc+1}))
  \end{math}

\end{center}

\subsubsection*{The final statement}

The problem gets to its end when the muddy children step forward.
This happens when \emph{muddy children know they are muddy}.  Formally
this is
\[ \forall i \in G,\quad \mud_i "->" K_i \mud_i\] Muddy children are able to
infer this statement when they know there are exactly $\nmc+1$ muddy
children: as every muddy child sees $\nmc$ ones (a persistent property),
he knows that he is muddy when he knows there are exactly
$\nmc+1$ muddy children, i.e. the $\nmc$ ones he sees plus him/herself.
\emph{If a child is muddy and if he knows there are exactly $\nmc+1$
  muddy children, then he knows he is muddy. }
This leads to the following axiom.
\[
\mathbf{MC3} \qquad \forall i`:G, \quad "|-" \mud_i "->" K_i\exact_{\nmc+1} "->" K_i \mud_i 
\]

\subsubsection*{The \pro~and the increase of knowledge}

The core of the work consists in clarifying formally what is produced
by Father's injunction and how this makes the muddy children's
knowledge to grow.

In this scenario, a tempo is given by  Father: time is made
discrete and is divided into time intervals which every agents (here
the children) can distinguish by counting Father's statements.
Therefore, these intervals can be numbered as follows:
\begin{itemize}
\item First interval starts at Father's declaration and ends at
  Father's first injunction
\item $(i+1)^{th}$ interval goes from $i^{th}$ to $(i+1)^{st}$
  injunction.
\end{itemize}

After $\nmc+1$ injunctions, every muddy child steps forward, as we will
prove it in our system for dynamic logic of common knowledge.  To do
so, we need to understand better what happens from an interval to
another with each Father's injunction.  These injunctions do not carry
much semantics, but they are important from a dynamic logic point of
view: indeed, each injunction gives a tempo and helps every child in
his quest of knowledge as it ends the previous interval.  Then every
child can deduce that no child has stepped forward during the previous
interval which means that none has been able to conclude about his
state, these increases the amount of information the children have..

Indeed, let us consider the first injunction. In the first interval,
$C_G\atleast_1$ holds and two cases occur:
\begin{description}
\item [If $\nmc=0$,] the only muddy child can say at once, that he is
  muddy because he is the only one to see no other muddy child and
  after Father's first injunction, he steps forward.
        
\item [If $\nmc>0$,] every child sees at least another muddy child, and
  so, no one can conclude whether he is muddy or not.  Worst,
  Father's initial statement of $\atleast_1$ did not tell them anything they
  do not know, but the fact that this statement became common
  knowledge and when no one steps forward at Father's first
  injunction, every child can infer that no one sees no muddy child,
  this means that every one sees at least one muddy child.  This can
  only happen if there are at least two muddy children. By an easy reasoning they exclude
  the case $\nmc=0$.
        
  To be more formal, \emph{every child knows that every child knows
    there is at least one muddy child}, which leads the children to
  the following: \emph{there are at least two muddy children}.
\end{description}

Father's first injunction translates formally into
\["|-" E_G E_G \atleast_1 "->" [*]E_G \neg \exact_1\]
which generalizes for any injunction:
\[
\mathbf{MC2} \qquad \forall j\in \{1,...,k\},\quad "|-" E_G E_G \atleast_j
"->" [*]E_G \neg \exact_j
\]
which is \emph{if every child knows that every child knows there are
  at least $j$ muddy children, then after Father's injunction, every
  child knows there cannot be exactly $j$ ones}.

\section{A knowledge gain lemma}
\label{sec:GainCom}

One can deduce a knowledge gain lemma which says that \emph{if
  every child knows that every child knows there are at least $j$
  muddy children, then after Father's injunction, every child knows
  there are at least $j+1$ ones}.  Formally

\begin{lemma}
  \(\mathbf{GainConn}\qquad \forall j\in \{1,...,k\},\quad "|-" E_G
  E_G \atleast_j "->" [*]E_G \atleast_{j+1} \)
\end{lemma}
\begin{proof}{~}
  Let $j\in \{1,...,k\}$.
  \begin{small}
    \[
    \prooftree %
    \prooftree %
    \prooftree %
    \prooftree %
    \prooftree \justifies "|-" E_G E_G\atleast_j "->" [*]E_G\neg \exact_j \using
    MC2
    \endprooftree
    \prooftree %
    \prooftree %
    \justifies "|-" E_GE_G\atleast_j "->" E_G\atleast_j \using T_E
    \endprooftree
    \prooftree %
    \prooftree \justifies "|-" \atleast_j "->" [*]\atleast_j \using PERS_{`v}
    \endprooftree
    \justifies "|-" E_G\atleast_j "->" [*]E_G\atleast_j \using EPers
    \endprooftree
    \justifies "|-" E_GE_G\atleast_j "->" [*]E_G\atleast_j \using Cut
    \endprooftree
    \justifies "|-" E_GE_G\atleast_j "->" [*]E_G\atleast_j \wedge [*]E_G\neg \exact_j
    \using \wedge Intro
    \endprooftree
    \justifies "|-" E_GE_G\atleast_j "->" [*](E_G\atleast_j \wedge E_G\neg \exact_j)
    \using */\wedge Dist
    \endprooftree
    \justifies "|-" E_GE_G\atleast_j "->" [*]E_G(\atleast_j \wedge \neg \exact_j) %
    \using E/\wedge Dist
    \endprooftree
    \justifies "|-" E_GE_G\atleast_j "->" [*]E_G \atleast_{j+1} \using IMP_{\atleast\exact}
    \endprooftree
    \]
  \end{small}
\end{proof}

\subsubsection*{Summary of the proof of the \mcp~theorem}

A common knowledge induces a nested shared knowledge at any level, the
{\bf GainConn} lemma deduced from {\bf MC2} axiom allows us to get a
picture of the proof of the \mcp~theorem, which we called {\bf Concl}
and which states as:
\[
\mathbf{Concl} \qquad "|-" \forall \nmc\in \mathbb N~~\forall i\in G,
\quad [\texttt{\currency}]\verb+TRUE+ "->" [*]^{\nmc} (\mud_i "->" K_i \mud_i)
\]

Indeed, initially, $\atleast_1$ is a common knowledge~({\bf MC1$_1$}), so it
is as an arbitrarily nested shared knowledge. With each Father's
injunction, children are able to make precise their knowledge about
the total number of muddy children by dropping one level of their
shared knowledge.  Therefore, after $j$ injunctions, they know
$\atleast_{j+1}$ by dropping $j$ depths of their shared knowledge. But since
initially this knowledge is arbitrarily deeply nested in shared
knowledge, after the first $\nmc$ Father's injunctions, every child
effectively knows $\atleast_{\nmc+1}$.

At this point, the muddy children know there are at least $\nmc+1$
muddy children; so, as they see $\nmc$ ones, they deduce there are
exactly $\nmc+1$ muddy children~({\bf MC1$_2$}) and they know they are
muddy themselves~({\bf MC3}). At the $(\nmc+1)^{st}$ injunction, they
will step forward miraculously, as Meyer and van der Hoek say with
humor.  After our \Coq{} experiments, we would say perfectly
logically!

One can notice that {\bf Concl} holds also for $\nmc=0$.  This theorem
describes all the scene: ``if Father makes its initial statement, then
after the $\nmc^{th}$ injunction, the agents who satisfy property $\mud$
know they do.''.


\section{The proof of the \mcp~theorem}

In this section, we describe the mechanized proof previously summed up
in more detail.

Let $\nc\in \mathbb N$ and $\nmc\in\{0,...,\nc+1\}$.

\begin{lemma}[MultGainConn] $\forall \nc\in \mathbb N^*~~\forall j\in
  \{0,...,\nmc\},\quad "|-" E_G^{\nc+1}\atleast_j "->" [*]E_G^{\nc} \atleast_{j+1}$
\end{lemma}               
\begin{proof}{~}
By induction on $\nc\in \mathbb N^*$ :
\begin{itemize}
\item
        Initialization : $\nc=1$,
        \[      \prooftree
                        \justifies "|-" E_G E_G\atleast_j "->" [*]E_G \atleast_{j+1} \using GainConn
                \endprooftree
        \]
\item Heredity : Let $\nc\in \mathbb N^*$,
        
  \begin{normalsize}
    \[ \prooftree \prooftree \prooftree \justifies "|-" E_G^{\nc+1}\atleast_j
    "->" [*]E_G^{n}\atleast_{j+1} \using {\footnotesize \texttt{HYP-REC}}
    \endprooftree
    \justifies "|-" E_GE_G^{\nc+1}\atleast_j "->" E_G[*]E_G^{n}\atleast_{j+1} \using
    EDist
    \endprooftree
    \prooftree \justifies "|-" E_G[*]E_G^{n}\atleast_{j+1} "->" [*]E_G^{\nc+1}
    \atleast_{j+1} \using KT1
    \endprooftree
    \justifies "|-" E_GE_G^{\nc+1}\atleast_j "->" [*]E_G^{\nc+1} \atleast_{j+1} \using
    Cut
    \endprooftree
    \]
  \end{normalsize}

\end{itemize}
\end{proof}

\begin{lemma}[ComImpPartIt] $\forall \nc\in \mathbb N,~~~~"|-" C_Gp "->" E_G^{\nc}p$
\end{lemma}
\begin{proof}{~}
By induction on $\nc\in \mathbb N$ :
\begin{itemize}
\item
        Initialization : $\nc=0$,
        \[      \prooftree
                        \prooftree \justifies "|-" C_Gp "->" p\wedge E_GC_Gp \using PointFixe_C \endprooftree
                        \justifies "|-" C_Gp "->" p \using \wedge Elim
                \endprooftree
        \]
\item Heredity : Let $\nc\in \mathbb N$,
        
\[      \prooftree
\prooftree 
\prooftree
\justifies "|-" C_Gp "->" p\wedge E_GC_Gp \using PointFixe_C
\endprooftree 
\justifies "|-" C_Gp "->" E_GC_Gp \using \wedge Elim
\endprooftree
\prooftree 
\prooftree
\justifies "|-" C_Gp "->" E_G^{n}p \using {\normalsize \texttt{HYP-REC}}
                                \endprooftree 
                                \justifies "|-" E_GC_Gp "->" E_G^{\nc+1}p \using E Dist
                        \endprooftree
                        \justifies "|-" C_Gp "->" E_G^{\nc+1}p \using Cut
                \endprooftree
        \]
\end{itemize}
\end{proof}

\begin{lemma}
  [PointImpPartIt] $\forall \nc\in \mathbb N^*,~~~~"|-"
  [\texttt{\currency}]\verb+TRUE+ "->" E_G^{\nc}\atleast_1$
\end{lemma}
\begin{proof}{~}
Let $\nc\in \mathbb N^*$.
        \[      \prooftree
                        \prooftree
                                \justifies "|-" [\texttt{\currency}]\verb+TRUE+ "->" C\atleast_1 \using MC1_1
                        \endprooftree
                        \prooftree
                                \justifies "|-" C_G\atleast_1 "->" E_G^{\nc}\atleast_1 \using ComImpPartIt
                        \endprooftree
                        \justifies "|-" [\texttt{\currency}]\verb+TRUE+ "->" E_G^{\nc}\atleast_1 \using Cut
                \endprooftree
        \]
\end{proof}

\begin{lemma}
  [PointImpProgr] $\forall \nc\geq k ~\forall j\in\{1,...,k+1\}, \\
 \indent \qquad  "|-"  [\texttt{\currency}]\verb+TRUE+ "->" [*]^{j-1}E_G^{\nc-j+1}\atleast_{j}$
\end{lemma}
\begin{proof}{~}
Let $\nc\geq k$.
                By induction on $j\in\{1,...,k+1\}$ :

\begin{itemize}
\item
        Initialization : $j=1$, 
        \[      \prooftree \justifies "|-" [\texttt{\currency}]\verb+TRUE+ "->" E_G^{\nc}\atleast_{1} \using PointImpPartIt \endprooftree \]
\item
Heredity : Let $j\in\{1,...,\nmc\}$,
\begin{footnotesize}
  \[ \prooftree \prooftree \prooftree \justifies "|-"
  [\texttt{\currency}]\verb+TRUE+ "->"
  [*]^{j-1}E_G^{\nc-j+1}\atleast_{j} \using {\scriptsize
    \texttt{HYP-REC}}
  \endprooftree
  \justifies "|-" [\texttt{\currency}]\verb+TRUE+ "->"
  [*]^{j-1}E_GE_G^{\nc-j}\atleast_{j} \using Id
  \endprooftree
  \prooftree \prooftree \justifies "|-" E_GE_G^{\nc-j}\atleast_{j} "->"
  [*]E_G^{\nc-j}\atleast_{j+1} \using {\scriptsize MultGainConn}
  \endprooftree
  \justifies "|-" [*]^{j-1}E_GE_G^{\nc-j}\atleast_{j} "->"
  [*]^{j}E_G^{\nc-j}\atleast_{j+1} \using (j-1)*Dist
  \endprooftree
  \justifies "|-" [\texttt{\currency}]\verb+TRUE+ "->"
  [*]^{j}E_G^{\nc-j}\atleast_{j+1} \using Cut
  \endprooftree
  \]
\end{footnotesize}

\end{itemize}
\end{proof}

From those lemma we get the following ones

\underline{With $j=\nmc+1$}
\begin{lemma}     
[ResInter$_1$]
$\forall \nc\geq \nmc,~~~~"|-" [\texttt{\currency}]\verb+TRUE+ "->"
[*]^{\nmc}E_G^{\nc-\nmc}\atleast_{\nmc+1} $
\end{lemma}

\underline{With $\nc=\nmc+1$}
\begin{lemma}
  [ResInter$_2$] $"|-" [\texttt{\currency}]\verb+TRUE+ "->" [*]^{\nmc}E_G\atleast_{\nmc+1}$
\end{lemma}
And the \mcp~theorem comes out (almost) easily.
\begin{theorem}[Concl]
  $"|-" \forall \nmc\in \mathbb N~~\forall i\in G, \quad
  [\texttt{\currency}]\verb+TRUE+ "->" [*]^{\nmc} (\mud_i "->" K_i \mud_i)$
\end{theorem}
\begin{proof}
  \[
  \prooftree \prooftree \prooftree \prooftree \prooftree \prooftree
  \prooftree \prooftree \prooftree \prooftree \prooftree \justifies
  "|-" [\texttt{\currency}]\verb+TRUE+ "->" [*]^{\nmc} E_G\atleast_{\nmc+1} \wedge ( \mud_i "->"
  K_i(\exact_{\nmc} \vee \exact_{\nmc+1})) \using ResInter_2 \& MC1_2
  \endprooftree
  \justifies "|-" [\texttt{\currency}]\verb+TRUE+ "->" [*]^{\nmc} E_G\atleast_{\nmc+1}
  \wedge[*]^{\nmc}(\mud_i "->" K_i(\exact_{\nmc} \vee \exact_{\nmc+1})) \using
  PERS_{MC1_2}
  \endprooftree
  \justifies "|-" [\texttt{\currency}]\verb+TRUE+ "->" [*]^{\nmc} ( E_G\atleast_{\nmc+1} \wedge
  ( \mud_i "->" K_i(\exact_{\nmc} \vee \exact_{\nmc+1}))) \using */\wedge Dist
  \endprooftree
  \justifies "|-" [\texttt{\currency}]\verb+TRUE+ "->" [*]^{\nmc} ( K_i\atleast_{\nmc+1} \wedge
  (\mud_i "->" K_i(\exact_{\nmc} \vee \exact_{\nmc+1}))) \using (E_Gp"->" K_ip)
  \endprooftree
  \justifies "|-" [\texttt{\currency}]\verb+TRUE+ "->" [*]^{\nmc} ( \mud_i "->"
  K_i\atleast_{\nmc+1} \wedge K_i(\exact_{\nmc} \vee \exact_{\nmc+1})) \using (a\wedge (b"->"
  c)"->" (b"->" a\wedge c))
  \endprooftree
  \justifies "|-" [\texttt{\currency}]\verb+TRUE+ "->" [*]^{\nmc} ( \mud_i "->"
  K_i(\atleast_{\nmc+1}\wedge (\exact_{\nmc} \vee \exact_{\nmc+1}))) \using K/\wedge Dist
  \endprooftree
  \justifies "|-" [\texttt{\currency}]\verb+TRUE+ "->" [*]^{\nmc} ( \mud_i "->"
  K_i((\atleast_{\nmc+1}\wedge \exact_{\nmc}) \vee (\atleast_{\nmc+1}\wedge \exact_{\nmc+1}))) \using
  \wedge /\vee Dist
  \endprooftree
  \justifies "|-" [\texttt{\currency}]\verb+TRUE+ "->" [*]^{\nmc} ( \mud_i "->"
  K_i((\atleast_{\nmc+1}\wedge \exact_{\nmc}) \vee\exact_{\nmc+1} )) \using (\atleast_{\nmc}\wedge
  \exact_{\nmc}"->" \exact_{\nmc+1})
  \endprooftree
  \justifies "|-" [\texttt{\currency}]\verb+TRUE+ "->" [*]^{\nmc} ( \mud_i "->" K_i(\bot
  \vee\exact_{\nmc+1} )) \using (EXCLU_{\atleast\exact})
  \endprooftree
  \justifies "|-" [\texttt{\currency}]\verb+TRUE+"->" [*]^{\nmc} ( \mud_i "->"
  K_i\exact_{\nmc+1} ) \using (\bot \vee p "->" p)
  \endprooftree
  \justifies "|-" [\texttt{\currency}]\verb+TRUE+"->" [*]^{\nmc} ( \mud_i "->" K_i \mud_i)
  \using MC2
  \endprooftree
  \]
\end{proof}

\section{The dynamic logic of common knowledge in \Coq}

\subsection{Implementation of $\mathcal T^{C[`a]}_G$ in \Coq}

\begin{sloppypar}
  See the appendix for few words on \Coq.  The implementation
  presented in this paper is based on another implementation, namely
  this of the Logic of Common Knowledge done by the first
  author~\cite{lescanne06:_mechan_coq_AMAI} who implemented all the
  epistemic multi-agent logic with common knowledge (system $\mathcal
  T^C_G$), of which a \Coq{} file is available on the web:
  \url{http://perso.ens-lyon.fr/pierre.lescanne/COQ/EpistemicLogic.v8}
\end{sloppypar}

This paper comes out with its own \Coq{} file:\\
\centerline{\small \url{http://perso.ens-lyon.fr/pierre.lescanne/COQ/EpistemicAndDynamicLogic.v}}
which implements the whole system $\mathcal T^{C[`a]}_G$ and a complete
proof of the \mcp~theorem {\bf Concl}.

\subsection{Why this implementation?}

The first aim of this implementation was to ensure a reader that the
proof is totally checkable.  This lead to a proof of nearly 1100 lines
of \Coq~code, where every lemma is the direct translation of the
hand-made proof for a maximal legibility.  We do not claim that proof
are readable as they would be in an English paper, a certain
technicality is required for giving all the detail of the proof; however
we claim that the statements of the lemmas are easily readable.

As an added value, this implementation allows any future development
by adding axioms or new modalities. This makes our work flexible and
reusable.


\section{Conclusion}

The proof theoretic approach we have used in this paper combines easily
epistemic and dynamic logics together, thanks to a general
epistemic-dynamic axiom ({\bf KT1}). \textbf{(KT1)} involves a
commutativity between epistemic modality and a dynamic modality.  In
the current implementation of \textbf{(KT1)}, type is not used to
check whether the axiom is only invoked on purely epistemic propositions.
In a future implementation, we will create a new type \emph{epistemic
  proposition} on which \textbf{(KT1)} can only be invoked.

After manipulating the logical system presented in this paper with the
proof assistant \Coq, we feel that it is quite simple and intuitive.
It only uses axioms and rules from classical logic plus a few
additional axioms and rules.  Statements can be made in a language
close to this of the hand proof.

The dynamic logic of common knowledge is based on knowledge and events.  In a formal statement, an
event becomes a dynamic modality which transforms a proposition that
describes the world before the event into a proposition that describes
the world after that event.  Said otherwise a dynamic modality
transforms properties into others. Here we have limited our work to
epistemic events which only transform agent knowledge, but this is not
a big restriction, as this is what happens most of the time.

We notice that we had to adapt the system for the specific situation
generated by the \mcp.  But this is no so different from situation where
classical logic or another system is used.  However, conceptual tools
or practical tools (for instance implemented in \Coq) could be built
to ease the task of the person who mechanizes a proof.

\paragraph*{Acknowledgments}
We would like to thank St\'{e}phane Le Roux, for discussion and advice
about \Coq.


\appendix
\section*{What is \Coq?}
\label{sec:WhatsCoq}

\Coq~is a proof assistant, i.e., a program which verifies step by step
the validity of a mathematical proof given by the user. In logic, it
is generally not obvious to follow a hand-made proof and to determine
whether it is right or wrong~\cite{weisstein:_kepler_conjec}. A proof
assistant, such as \Coq, becomes a necessary tool if one chooses to be
absolutely sure of a result.

Moreover, \Coq~is a very good means to build proofs.  Indeed, managing
a proof step by step, as required by a proof assistant, allows us to
understand in a very precise way what is done and what has to be
done to complete a proof. \Coq~is also a way to reach a good formalism
as it requires from the user to define exactly all what he
manipulates.  

\section*{Excerpts of the Coq script}

Here is the statement of the main lemmas and of the last theorem
\textbf{Concl}.

\begin{footnotesize}
\begin{verbatim}
Lemma GainConn :
  forall  (G: list nat) (i j : nat), 
    |- E (i::G) (E (i::G) (lambda j)) ==> 
        [*] (E (i::G) (lambda (j+1))).

Lemma MultGainConn :
  forall (G: list nat) (m i j : nat), 
    |- F ((m+1)+1) (i::G) (lambda j) ==> 
        [*] (F (m+1) (i::G) (lambda (j+1))).

Lemma ComImpPartIt : 
  forall (p:proposition) (n:nat) (G: list nat), 
    |- C G p ==> F n G p.

Lemma PointImpPartIt : 
  forall (G:list nat) (m:nat), 
    |- [¤] TRUE ==> F m G (lambda 1).

Lemma PointImpProgr : 
  forall (G:list nat) (i j n:nat), 
    |- [¤] TRUE ==> [*]<:j:> (F (n+1) (i::G) (lambda (j+1) ) ).

Lemma Concl : 
  forall (G:list nat) (i j m : nat), In i (j::G) -> 
    |- [¤] TRUE ==> [*]<:m:> (muddy i ==> (K i (muddy i) ) ).

\end{verbatim}
\end{footnotesize}
\end{document}